\documentclass[twocolumn,showpacs,prd]{revtex4}
\usepackage{mathrsfs}
\usepackage{longtable,lscape}
\usepackage{txfonts}
\usepackage{amssymb}
\usepackage{indentfirst}
\usepackage{graphicx,,booktabs}
\usepackage{color}
\usepackage{amssymb}

\begin{document}

\title{Anomalous dipion invariant mass distribution of the
$\Upsilon(4S)$ decays into $\Upsilon(1S) \pi^{+} \pi^{-}$ and $\Upsilon(2S) \pi^{+} \pi^{-}$}
\author{Dian-Yong Chen$^{1,3}$}
\author{Xiang Liu$^{1,2}$\footnote{Corresponding author}}\email{xiangliu@lzu.edu.cn}

\affiliation{
$^1$Research Center for Hadron and CSR Physics,
Lanzhou University and Institute of Modern Physics of CAS, Lanzhou 730000, China\\
$^2$School of Physical Science and Technology, Lanzhou University, Lanzhou 730000,  China\\
$^3$Nuclear Theory Group, Institute of Modern Physics of CAS, Lanzhou 730000, China}
\author{Xue-Qian Li}
\affiliation{School of Physics, Nankai University, Tianjin 300071,
China}
\date{\today}

\begin{abstract}
To solve the discrepancy between the experimental data on the partial widths and lineshapes of the dipion emission of
$\Upsilon(4S)$ and the theoretical predictions, we suggest that there is an additional contribution, which was not taken into account
in previous calculations. Noticing that the mass of $\Upsilon(4S)$ is above the production threshold of $B\bar B$, the contribution of the sequential process
$\Upsilon(4S)\to B\bar B\to \Upsilon(nS)+S\to \Upsilon(nS)+\pi^+\pi^-$ ($n=1,2$) may be sizable, and its interference with that from the direct production would be important.
The goal of this work is to investigate if a sum of the two contributions with a relative phase indeed
reproduces the data. Our numerical results on the partial widths and the lineshapes $d\Gamma(\Upsilon(4S)\to \Upsilon(2S,1S)\pi^+\pi^-)/d(m_{\pi^+\pi^-})$
are satisfactorily consistent with the measurements, thus the role of this mechanism is confirmed. Moreover, with the parameters obtained by fitting the data of the Belle and Babar collaborations,
we predict the distributions $d\Gamma(\Upsilon(4S)\to \Upsilon(2S,1S)\pi^+\pi^-)/d\cos\theta$, which have not been measured yet.

\end{abstract}

 \maketitle

The dipion emissions of the bottomonia were observed a long while ago, and as an important channel
to investigate the decay mechanism and inner structure of bottomonia, they have undergone careful studies
from both theoretical and experimental sides.  The significance of the transitions stimulated
a series of theoretical efforts \cite{Yan:1980uh,Kuang:1981se,Zhou:1990ik,Guo:2006ai,Simonov:2008qy}
(see Refs. \cite{Voloshin:1987rp,Besson:1993mm,Kuang:2006me} for a review). Before $\Upsilon(4S)$ joined the game,
the transitions occurring among only the bottomonia $\Upsilon(3S)$ and $\Upsilon(1S,2S)$ were
below the $B\bar{B}$ threshold \cite{Nakamura:2010zzi}. With  $\Upsilon(4S)$ being experimentally observed,
the situation becomes more intriguing because the $B\bar{B}$ channel is open. It is similar to the case of $\psi(3770)$,
which is a charmonium with a mass just above the $D\bar{D}$ threshold. The CLEO-c Collaboration  explored the dipion emission
of $\Upsilon(4S)$, however, only upper limits for $\Upsilon(4S)\to \Upsilon(1S)\pi^+\pi^-$ and
$\Upsilon(4S)\to \Upsilon(2S)\pi^+\pi^-$ were obtained \cite{Glenn:1998bd}. With the data accumulation of the two $B$-factories,
more precise measurement on  the dipion emission of $\Upsilon(4S)$ became possible.

In recent years, a great progress on measurements of the $\Upsilon(4S)$ dipion transitions has been made
\cite{Abe:2005bi,Sokolov:2006sd,Aubert:2006bm}. The Belle Collaboration reported their observation of
$\Upsilon(4S)\to \Upsilon(1S)\pi^+\pi^-$ \cite{Abe:2005bi,Sokolov:2006sd} whose partial
decay width is $\Gamma(\Upsilon(4S)\to \Upsilon(1S)\pi^+\pi^-)=(3.65\pm0.67(\mathrm{stat})\pm0.65(\mathrm{syst}))$ keV.
The BaBar Collaboration also presented their measurements on $\Upsilon(4S)$ decays into $\Upsilon(1S)\pi^+\pi^-$ and $\Upsilon(2S)\pi^+\pi^-$ \cite{Aubert:2006bm},
and the partial widths are respectively $\Gamma(\Upsilon(4S)\to \Upsilon(1S)\pi^+\pi^-)=(1.8\pm0.4)$ keV and
$\Gamma(\Upsilon(4S)\to \Upsilon(2S)\pi^+\pi^-)=(2.7\pm0.8)$ keV \cite{Aubert:2006bm}.
Besides measuring the decay rates of $\Upsilon(4S)\to \Upsilon(1S)\pi^+\pi^-$ and $\Upsilon(4S)\to \Upsilon(2S)\pi^+\pi^-$,
the Belle and BaBar collaborations also obtained the  dipion invariant mass distribution, which is defined as $d\Gamma(\Upsilon(4S)\to \Upsilon(2S,1S)\pi^+\pi^-)/d(m_{\pi^+\pi^-})$
versus $m_{\pi^+\pi^-}$ and $d\Gamma(\Upsilon(4S)\to \Upsilon(2S,1S)\pi^+\pi^-)/d\cos\theta$ versus $\cos\theta$, where $\theta$ is the angle between $\Upsilon(4S)$ and $\pi^-$ in
the $\pi^+\pi^-$ rest frame.

As shown in Fig. 3 of Ref. \cite{Aubert:2006bm},  a comparison between the experimental data
and theoretical prediction \cite{Kuang:1981se} on the $m_{\pi^+\pi^-}$ distributions for $\Upsilon(4S)\to \Upsilon(1S)\pi^+\pi^-$
and $\Upsilon(4S)\to \Upsilon(2S)\pi^+\pi^-$ decays was given. The measured $m_{\pi^+\pi^-}$ distribution for
$\Upsilon(4S)\to \Upsilon(1S)\pi^+\pi^-$ is close to that predicted in Ref. \cite{Kuang:1981se},
whereas the measured lineshape of the $m_{\pi^+\pi^-}$ distribution for the $\Upsilon(4S)\to \Upsilon(2S)\pi^+\pi^-$
fully deviates from the prediction of the Kuang-Yan model \cite{Kuang:1981se}. Considering that the model is completely successful in describing
heavy quarkonium transitions $\Upsilon(2S)\to \Upsilon(1S)\pi^+\pi^-$, $\Upsilon(3S)\to \Upsilon(1S)\pi^+\pi^-$ and $\psi(2S)\to J/\psi\pi^+\pi^-$, one
should conclude that there must be some mechanism, which has not been taken into account in the old calculations.

Moreover, the ratio ${\mathcal{R}}=\frac{\Gamma(\Upsilon(4S)\to \Upsilon(2S)\pi^+\pi^-)}{\Upsilon(4S)\to \Gamma(\Upsilon(1S)\pi^+\pi^-)}\approx1.16$
was given by the BaBar Collaboration \cite{Aubert:2006bm}. Generally speaking, the $\Upsilon(1S)\pi^+\pi^-$ channel possesses a larger
phase space than that of $\Upsilon(2S)\pi^+\pi^-$, and it should result in a ratio $\mathcal{R}$ to be
smaller than unity. As a good example, the rule is well satisfied for $\Upsilon(3S)$ as the measured $\mathcal{R}=0.577$
\cite{:2008bv}. The experimental value of the ratio ${R}$ for $\Upsilon(4S)$ obviously declines from the
rule, which might also imply existence of some uncounted factors. The very recent experimental result on the ratio $\mathcal{R}$ of $\Upsilon(5S)$ is given by the Belle Collaboration as $R=1.44$ \cite{Abe:2007tk},
which also contradicts to the simple argument. Thus, motivated by these phenomena, one should conjecture that
anomalous production rates and lineshapes of $\Upsilon(5S,4S)\to \Upsilon(1S)\pi^+\pi^-,\Upsilon(2S)\pi^+\pi^-$ are due to the available open $B\bar B$ channels \cite{Chen:2011qx,Chen:2011zv}.

$\Upsilon(4S)$ is obviously different from the lower states. First, the mass of $\Upsilon(4S)$ is just above the $B\bar{B}$ threshold,
so that the coupled channel effect becomes important. Another feature is that $\Upsilon(4S)$ predominantly decays into $B\bar{B}$ pair because the mode is
not suppressed by the OZI rule.
Thus, when exploring  hidden-bottom decays of $\Upsilon(4S)$, we cannot ignore
the coupled-channel effect, which may play a crucial role in fact.

\begin{figure}[htb]
\centering
\begin{tabular}{ccc}
\scalebox{0.9}{\includegraphics{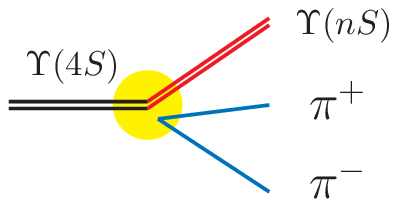}}&\raisebox{4.8ex}{\huge{$+$}}&
\scalebox{0.9}{\includegraphics{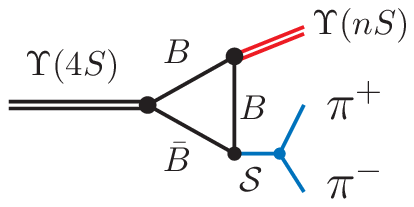}}\\
(a)&&(b)
\end{tabular}
\caption{The diagrams relevant to $\Upsilon(4S)$ hidden-bottom decay with dipion emission.
Here, Fig. \ref{decay} (a) represents $\Upsilon(4S)$ direct decay into $\Upsilon(nS)\pi^+\pi^-$ ($n=1,2$), while
Fig. \ref{decay} (b) denotes the intermediate hadronic loop contribution to $\Upsilon(4S)\to \Upsilon(nS)\pi^+\pi^-$.}
\label{decay}
\end{figure}

In general, there exist two mechanisms for the dipion emission
$$\Upsilon(4S)\to
\Upsilon(nS)(p_1)\pi^+(p_2)\pi^-(p_3)\quad \mathrm{with}\quad(n=1,2).$$ One is a direct process, where two gluons are successively emitted and eventually
hadronized into two pions,
which is depicted in Fig. \ref{decay} (a) and can be described by the QCD Multipole Expansion (QME) model \cite{Kuang:1981se}. Applications of the method
have been discussed in all details, thus
in this letter, we do not repeat the calculation for the direct transition,
but alternatively introduce an effective Lagrangian to take care of this contribution.

The transition amplitude of the direct transition $\Upsilon(4S)\to\Upsilon(nS)\pi^+\pi^-$ can be written as
\begin{eqnarray}
&&\mathcal{M}[\Upsilon(4S)\to
\Upsilon(nS)\pi^+\pi^-]_{\mathrm{Direct}}\nonumber\\&&
={\mathcal{F}^{(n)}\over{f_\pi^2}}\epsilon_{\Upsilon(5S)}\cdot
\epsilon_{\Upsilon(nS)}\Big\{\Big[q^2-\kappa^{(n)}(\Delta
M)^2\Big(1+\frac{2m^2_\pi}{q^2}\Big)\Big]_{\mathrm{S-wave}}\nonumber\\&&
\quad+\Big[\frac{3}{2}\kappa^{(n)}\big((\Delta M)^2-q^2\big)
\Big(1-\frac{4m_\pi^2}{q^2}\Big)
\Big(\cos\theta^2-\frac{1}{3}\Big)\Big]_{\mathrm{D-wave}}\Big\}
,\nonumber\\\label{direct}
\end{eqnarray}
which was suggested by Novikov and Shifman in studying $\psi^\prime\to J/\psi\pi^+\pi^-$ \cite{Novikov:1980fa} and the subscripts (S-wave) and (D-wave)
denote the S-wave and D-wave contributions respectively. The mass
difference between $\Upsilon(4S)$ and $\Upsilon(nS)$ is expressed as $\Delta M$.
$q^2=(p_2+p_3)^2\equiv m_{\pi^+\pi^-}^2$ denotes the invariant mass of
$\pi^+\pi^-$, while $\theta$ is the angle between $\Upsilon(4S)$ and
$\pi^-$ in the $\pi^+\pi^-$ rest frame. The pion decay constant and mass are taken as $f_\pi=130$ MeV and $m_\pi=140$ MeV, respectively.
In Eq. (\ref{direct}), $\kappa$ and $\mathcal{F}$ are left as free parameters when fitting the experimental data.

There also exists another mechanism shown in Fig. \ref{decay} (b)
contributing to the $\Upsilon(4S)\to \Upsilon(nS)\pi^+\pi^-$ transitions,
where the intermediate hadronic loop constructed by $B$ and $\bar{B}$ mesons
performs as a bridge to connect the initial state $\Upsilon(4S)$ and final state $\Upsilon(nS)\pi^+\pi^-$.
This manifests the coupled channel effect, an important non-perturbative QCD effect.
In other words, $\Upsilon(4S)\to \Upsilon(nS)\pi^+\pi^-$ can be approximately expressed as a sequential
decay process. $\Upsilon(4S)$ first transits into $\Upsilon(nS)$ and scalar state $\mathcal{S}$, then
$\mathcal{S}$ couples with dipion and the picture is depicted in the right panel of Fig. \ref{decay}.
For $\Upsilon(4S)\to \Upsilon(1S)\pi^+\pi^-$ and $\Upsilon(4S)\to \Upsilon(2S)\pi^+\pi^-$
processes, $\mathcal{S}$ are $\sigma(600),f_0(980)$ and $\mathcal{S}=\{\sigma(600)$
respectively, which are of proper quantum numbers and allowed by the phase space of the corresponding modes.
Since the intermediate $B$ and $\bar{B}$ mesons can be on-shell, as usually considered, the contribution from the absorptive part
of the loop is larger than the dispersive part, thus let us just keep the imaginary part of the loop. This definitely brings up errors, but
in our strategy, we fit the data to obtain the model parameters, so that part of the uncertainty would be compensated.
By the Cutosky cutting rule, the transition amplitude
corresponding to the absorptive part of the hadronic loop is written as
\begin{eqnarray}
&&\mathcal{M}[\Upsilon(4S)\to B\bar{B} \to
\Upsilon(nS)\mathcal{S}\to\Upsilon(nS)\pi^+\pi^-]_{\mathrm{FSI}}^{(\mathcal{S})} \nonumber\\
&&= \xi\frac{|\mathbf{k}|}{32\pi^2 M_{\Upsilon(4S)}}\int d\Omega
\mathcal{A}^*[ \Upsilon(4S)\to
B^+(k_1)B^-(k_2)]\nonumber\\&&\quad\times\mathcal{A}[B^+(k_1)B^-(k_2)\to
\Upsilon(nS)(p_1)\mathcal{S}(q)]\frac{g_{_{\mathcal{S}\pi\pi}}}{q^2-m_{\mathcal{S}}^2+im_S\Gamma_{\mathcal{S}}}\nonumber\\
&&\equiv\big[g^{(n)}_{0S} g_{\mu\nu}(p_1\cdot
q)+g_{0D}^{(n)}p_{1\mu}q_\nu\big]\frac{\epsilon_{\Upsilon(4S)}^\mu
\epsilon_{\Upsilon(nS)}^\nu}{q^2-m_S^2+i m_S \Gamma_S } g_{_{S
\pi\pi}} p_2 \cdot p_3,\nonumber\\\label{para}
\end{eqnarray}
where the factor $\xi=4$ is due to charge conjugation
$B\rightleftharpoons\bar{B}$ and isospin transformation
$B^0\rightleftharpoons B^+$ and $\bar{B}^0\rightleftharpoons B^-$.
Using the effective Lagrangian approach, we get
$\mathcal{A}^*[\Upsilon(4S)\to
B^+B^-]=ig_{\Upsilon(4S)BB}\epsilon_{\Upsilon(4S)}^\mu(ik_{2\mu}-ik_{1\mu})$
and $\mathcal{A}[B^+B^-\to
\Upsilon(nS)\mathcal{S}]=ig_{\Upsilon(nS)BB}\epsilon_{\Upsilon(nS)}^\rho
(-ik_{1\rho}-ip_\rho)\mathcal{F}^2[m_B^2,p^2]$ with the monopole
form factor
$\mathcal{F}[m_B^2,p^2]=(\Lambda^2-m_B^2)/(\Lambda^2-p^2)$, where
$m_B$ is the mass of the exchanged meson. The variable $\Lambda$
is usually re-parameterized as $\Lambda= m_B + \alpha
\Lambda_{QCD}$ with $\Lambda_{QCD}=220$ MeV. This monopole form
factor is introduced not only to reflect the structure of the
effective vertex for the $B^+{B}^-\to \Upsilon(nS)\mathcal{S}$
scattering process, but also concerns the off-shell effect of the
the exchanged $B$-mesons at the t-channel. The coupling constants
for the $\Upsilon(4S) BB$ is estimated by fitting the partial
decay width and $g_{\Upsilon(4S) BB} \simeq 24$, while for
$\Upsilon(nS)B B$ $(n=1,2)$ and $\mathcal{S} B B$, the coupling
constants are directly taken from Ref. \cite{Meng:2007tk}. In
addition, the coupling constants between the intermediate scalar
meson and the final $\pi^+ \pi^-$ are $g_{\sigma \pi \pi} =16.2$
GeV$^{-1}$ and $g_{f_0 \pi \pi}=2.40$ GeV$^{-1}$, which are
determined by fitting the corresponding partial widths
\cite{Nakamura:2010zzi,Aitala:2000xu}. Finally we parameterized
the whole hadronic loop contribution as a simple and explicit
Lorentz structure presented in the last expression of Eq.
(\ref{para}), where $g_{0S}^{(n)}$ and $g_{0D}^{(n)}$,
corresponding to S-wave and D-wave contribution respectively, are
extracted from the hadronic loop calculation and dependent on the
dipion invariant mass $m_{\pi^+\pi^-}$.

With the above preparation illustrated above, the total transition amplitude
of $\Upsilon(4S)\to \Upsilon(nS)\pi^+\pi^-$ is expressed as
\begin{eqnarray}
\mathcal{M}_{\mathrm{Total}}=\mathcal{M}_{\mathrm{Direct}}+
\sum_\mathcal{S} e^{i\phi_\mathcal{S}^{(n)}}\mathcal{M}_{\mathrm{FSI}}^{(\mathcal{S})},\label{in}
\end{eqnarray}
where the phase $\phi_\mathcal{S}^{(n)}$ is introduced by hand to account for a possible relative phase between the direct transition and
the hadronic loop contributions. As a three-body decay, the
differential decay width for $\Upsilon(4S) \to \Upsilon(nS) \pi^+
\pi^-$ reads as,
\begin{eqnarray}
d\Gamma = \frac{1}{(2 \pi)^3} \frac{1}{32
m_{\Upsilon(4S)}^3} \overline{|\mathcal{M}_{\mathrm{total}}|^2}
dm_{\Upsilon(nS) \pi}^2 dm_{\pi^+\pi^-}^2
\end{eqnarray}
with $m_{\Upsilon(nS) \pi^+}^2 = (p_1 + p_2)^2$ and $m_{\pi^+\pi^-}^2 =(p_2
+p_3)^2$. The over-lined bar indicates an average over the polarizations
of $\Upsilon(4S)$ in the initial state and sum over the
polarizations of $\Upsilon(nS)$ in the final state. The parameters about the concerned resonances are listed in Table. \ref{Tab-Input}.
\begin{table}[htb]
\centering %
\caption{The resonance parameters adopted in this
work \cite{Nakamura:2010zzi,Aitala:2000xu}. \label{Tab-Input}}
\begin{tabular}{cccccc}
 \toprule[1pt]
&Mass (GeV)&&Mass (GeV)&Width (GeV)\\\midrule[1pt]
$\Upsilon(1S)$&9.469&$\sigma(600)$&0.526&0.302\\
$\Upsilon(2S)$&10.024&$f_0(980)$&0.98&0.07\\
$\Upsilon(4S)$&10.579&&&\\
 \bottomrule[1pt]
\end{tabular}
\end{table}

In our present scenario, we keep four parameters
$\mathcal{F}^{(1)}$, $\kappa^{(1)}$, $\phi_\sigma^{(1)}$,
$\phi_{f_0(980)}^{(1)}$ free for $\Upsilon(4S)\to
\Upsilon(1S)\pi^+\pi^-$  and three free parameters
$\mathcal{F}^{(2)}$, $\kappa^{(2)}$, $\phi_\sigma^{(2)}$ for
$\Upsilon(4S)\to \Upsilon(2S)\pi^+\pi^-$. So far, there are three
measurements on the dipion invariant mass spectrum  of
$\Upsilon(4S)\to \Upsilon(1S)\pi^+\pi^-$ given by Belle
\cite{Abe:2005bi,Sokolov:2006sd} and BaBar \cite{Aubert:2006bm}
also measured the $m_{\pi^+\pi^-}$ distribution of
$\Upsilon(4S)\to\Upsilon(2S)\pi^+\pi^-$. Our strategy is
following. We have the Belle and BaBar data on the partial widths
of $\Upsilon(4S)\to\Upsilon(2S)\pi^+\pi^-$ and
$\Upsilon(4S)\to\Upsilon(1S)\pi^+\pi^-$, and the measure dipion
distributions of $d\Gamma/d m_{\pi^+\pi^-}$, then using our model
we make a best fit to the lineshapes when keeping the partial
width to be consistent with the measured values. Namely, we will
check whether the Belle  \cite{Abe:2005bi,Sokolov:2006sd} and
BaBar data \cite{Aubert:2006bm} can be reproduced by our model.
Our numerical computation is done with the help of the MINUIT
package.

\begin{figure}[htb]
\begin{center}
\scalebox{0.575}{\includegraphics{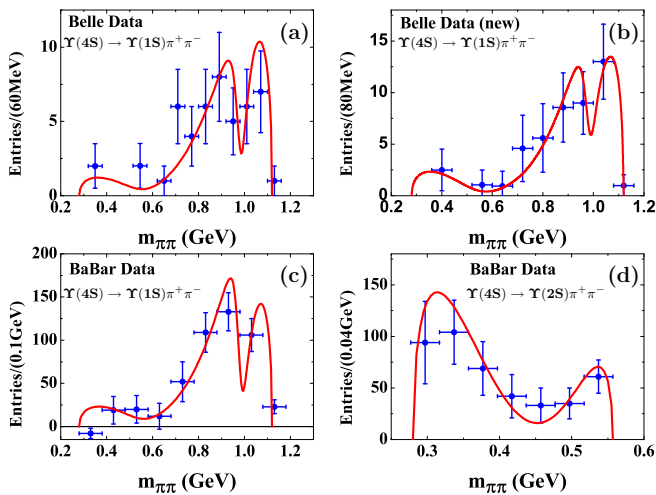}}%
\caption{The dipion invariant mass distributions for
$\Upsilon(4S) \to \Upsilon(1S) \pi^+ \pi^-$ and $\Upsilon(4S) \to \Upsilon(2S) \pi^+ \pi^-$
transitions. Here, the red solid curves
are the best fit to the experimental data (blue points with error
bars) \cite{Abe:2005bi,Sokolov:2006sd,Aubert:2006bm} in our model. Here,
the data points with error bars in (a) and (b) correspond to the measured results given by Belle
in Ref. \cite{Abe:2005bi} and a published work \cite{Sokolov:2006sd}, respectively. \label{4s-1s}}
\end{center}
\end{figure}

As shown in Fig. \ref{4s-1s}, we carry out the best fit (red solid curves) to the experimental data of
$\Upsilon(4S)\to\Upsilon(1S)\pi^+\pi^-$ and $\Upsilon(4S)\to\Upsilon(2S)\pi^+\pi^-$. The central values along
with the errors of the fitted parameters are listed in Tables. \ref{1s} and \ref{2s}.

\begin{figure}[htb]
\begin{center}
\scalebox{0.68}{\includegraphics{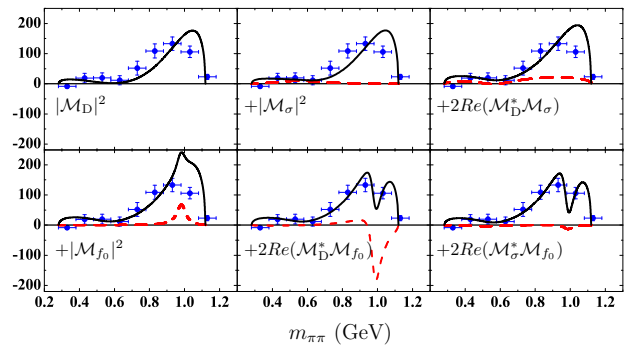}}%
\caption{As an example, we illustrate the changes of the line shape (black solid lines) for  fitting the BaBar data (blue dots with error) of $\Upsilon(4S)\to\Upsilon(1S)\pi^+\pi^-$ \cite{Abe:2005bi}
by adding the contributions from all the six modulus-square of individual amplitudes
($|\mathcal{M}_{\mathrm{D}}|^2$, $|\mathcal{M}_\sigma|^2$,
$2Re(|\mathcal{M}_{\mathrm{D}}^*\mathcal{M}_\sigma)$,
$|\mathcal{M}_{f_0}|^2$,
$2Re(\mathcal{M}_{\mathrm{D}}^*\mathcal{M}_{f_0})$ and $2Re(\mathcal{M}_{\sigma}^*\mathcal{M}_{f_0})$) one by one. Here, the red dashed lines
correspond to these modulus-square of the individual amplitudes.  \label{step}}
\end{center}
\end{figure}

For illustrating how to fit the line shape of $\Upsilon(4S)\to
\Upsilon(1S,2S)\pi^+\pi^-$ step by step, let us provide some details
about the changes of theoretically predicted line shapes (see Fig.
\ref{step}) of $\Upsilon(4S)\to \Upsilon(1S)\pi^+\pi^-$ while
considering the interference shown in Eq. (\ref{in}). Since both
$\sigma(600)$ and $f_0(980)$ contribute to Fig. \ref{decay} (b),
$|\mathcal{M}_{\mathrm{Total}}|^2$ can be decomposed into six terms
$|\mathcal{M}_{\mathrm{D}}|^2$, $|\mathcal{M}_\sigma|^2$,
$2Re(|\mathcal{M}_{\mathrm{D}}^*\mathcal{M}_\sigma)$,
$|\mathcal{M}_{f_0}|^2$,
$2Re(\mathcal{M}_{\mathrm{D}}^*\mathcal{M}_{f_0})$ and $2Re(\mathcal{M}_{\sigma}^*\mathcal{M}_{f_0})$, where we
abbreviate
$\mathcal{M}_{\mathrm{Direct}}\to\mathcal{M}_{\mathrm{D}}$,
$\mathcal{M}_{\mathrm{FSI}}^{(\sigma)}\to \mathcal{M}_\sigma$ and
$\mathcal{M}_{\mathrm{FSI}}^{(f_0)}\to \mathcal{M}_{f_0}$.

It is noted that for presenting the dipion invariant mass distribution, the experimental
colleagues use a different quantity "entries/mass", which
exactly corresponds to the distribution $d\Gamma(\Upsilon(4S)\to \Upsilon(2S,1S)\pi^+\pi^-)/d(m_{\pi^+\pi^-})$,
where the different lines correspond to the parameters obtained by fitting different data bases as indicated in the figures.
Thus when we plot the dependence of the partial width over $m_{\pi^+\pi^-}$, we
deliberately normalize our numbers for a real comparison with the available data.

\begin{table}[htbp]
\centering%
\caption{The values of parameters for the best fit to the experimental
invariant mass distributions of $\Upsilon(4S) \to \Upsilon(1S) \pi^+ \pi^-$. Here, the
values listed in
the second,  third and  fourth columns correspond to the best fits shown in diagrams (a), (b) and (c) of Fig. \ref{4s-1s},
respectively. The last row lists the partial decay
width for $\Upsilon(4S) \to \Upsilon(1S) \pi^+ \pi^-$ obtained
by using the central values of those parameters.\label{1s} }
\begin{tabular}{cccccc}
 \toprule[1pt]
 & Belle & Belle (new) & BaBar \\
 \midrule[1pt]
 $\mathcal{F}^{(1)}$     &    $0.045 \pm   0.002$  &  $0.063 \pm 0.003$ & $ 0.040 \pm 0.001$ \\
 $\kappa^{(n)}$          &    $0.249 \pm   0.041$  &  $0.311 \pm 0.018$ & $ 0.269 \pm 0.023$ \\
 $\phi_\sigma^{(1)}$     &    $3.124 \pm   1.156$  &  $-2.64 \pm 0.609$ & $ -3.14 \pm 0.324$ \\
 $\phi_{f_0(980)}^{(1)}$ &    $1.207 \pm   0.489$  &  $0.973 \pm 0.723$ & $ 0.918 \pm 0.183$ \\
$\Gamma$(keV)     & $2.255$           & $3.690$            & $1.799$            \\
 \bottomrule[1pt]
\end{tabular}
\end{table}

\begin{table}[htbp]
\centering%
\caption{The parameters obtained from the best fit to the dipion
invariant mass distribution of $\Upsilon(4S) \to \Upsilon(2S) \pi^+ \pi^-$. The last row gives the partial decay
width for $\Upsilon(4S) \to \Upsilon(2S) \pi^+ \pi^-$ obtained in our model. \label{2s}}
\begin{tabular}{cccccc}
 \toprule[1pt]
$\mathcal{F}^{(2)}$      & $0.838 \pm 0.042$ &
$\kappa^{(2)}$           & $0.664 \pm 0.029$ \\
$\phi_\sigma^{(2)}$     & $-2.923 \pm 0.141$ &
$\Gamma$(keV)     & $2.642$\\
 \bottomrule[1pt]
\end{tabular}
\end{table}

\begin{figure}[htb]
\centering%
\scalebox{0.35}{\includegraphics{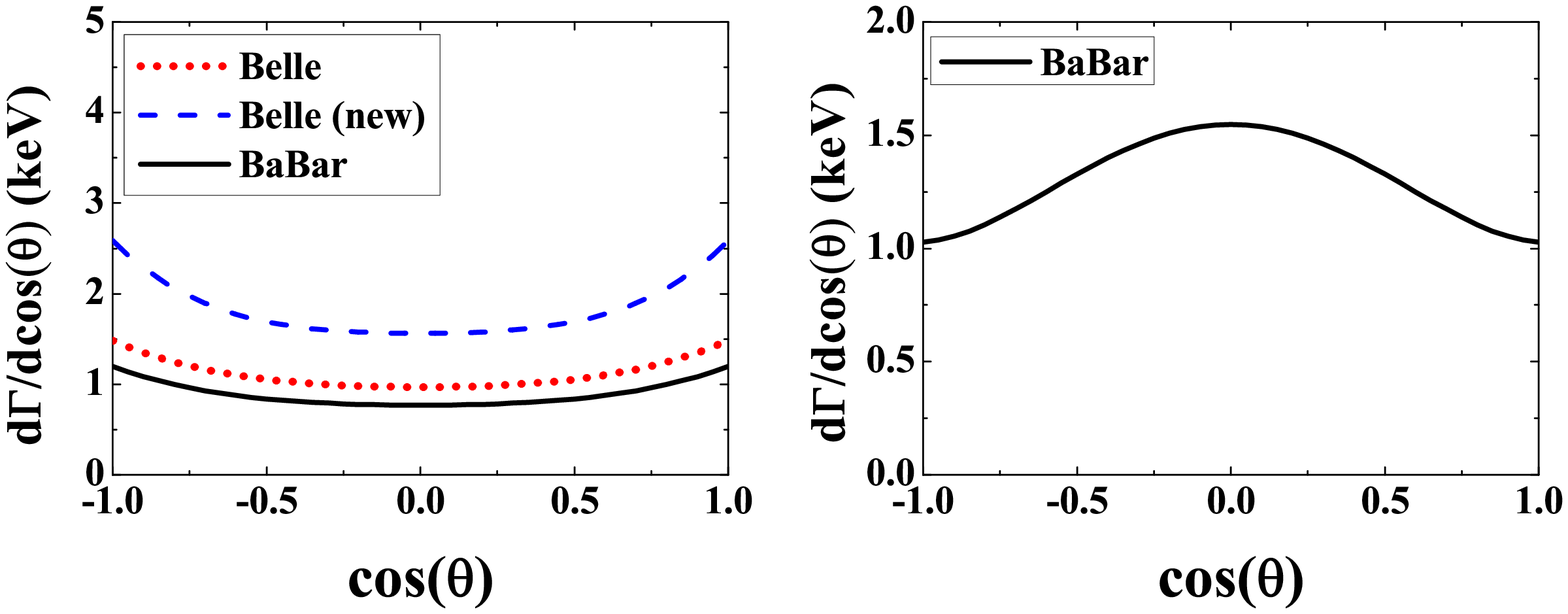}}%
\caption{The predicted $d\Gamma/d\cos\theta$ of $\Upsilon(4S) \to \Upsilon(1S) \pi^+ \pi^-$ (left panel) and
$\Upsilon(4S)\to \Upsilon(2S) \pi^+ \pi^-$ (right panel)
transitions with the obtained values of the fitting parameters listed in Tables. \ref{1s}-\ref{2s}. \label{haha}}
\end{figure}

As shown in Fig. \ref{4s-1s}, the Belle data presented in Ref.
\cite{Abe:2005bi} indicate that a double-peak structure appears in
the range of $m_{\pi^+\pi^-}=0.8\sim 1.2$ GeV, while later measured
results given in Refs. \cite{Sokolov:2006sd,Aubert:2006bm} do not
show an obvious double-peak structure. This difference may be caused
by different bin widths of the experimental settings of the Belle
\cite{Abe:2005bi,Sokolov:2006sd} and BaBar \cite{Aubert:2006bm}.
Thus, in this work we separately fit their different experimental
data. Further experimental measurement will test whether there exist
this double-peak structure in the dipion invariant mass spectrum of
$\Upsilon(4S) \to \Upsilon(1S) \pi^+ \pi^-$.

In addition, we notice that the experimental data of $\Upsilon(4S)
\to \Upsilon(1S) \pi^+ \pi^-$ and $\Upsilon(4S) \to \Upsilon(2S)
\pi^+ \pi^-$ are quite apart from each other. Our analysis indicates
that considering background and intermediate $\sigma(600)$ and
$f_0(980)$ contributions to $\Upsilon(4S) \to \Upsilon(1S) \pi^+
\pi^-$, both of the experimental data
\cite{Abe:2005bi,Sokolov:2006sd,Aubert:2006bm} can be well met. For
$\Upsilon(4S) \to \Upsilon(2S) \pi^+ \pi^-$ process, there exist
both background and intermediate $\sigma(600)$ contribution, and it
is different from the case of $\Upsilon(4S) \to \Upsilon(1S) \pi^+
\pi^-$. Thus, the interference between background and intermediate
$\sigma(600)$ contributions reasonably results in a smooth line
shape of the distribution of the dipion invariant mass spectrum of
$\Upsilon(4S) \to \Upsilon(2S) \pi^+ \pi^-$.

It is noted that with the parameters obtained by fitting the
lineshapes of the Belle and BaBar collaborations on
$D\Gamma/dm_{\pi^+\pi^-}$ for $\Upsilon(4S) \to \Upsilon(1S) \pi^+
\pi^-$ and $\Upsilon(4S) \to \Upsilon(2S) \pi^+ \pi^-$, we predict
the angular distribution $d\Gamma/d\cos\theta$ as shown in Fig. 3.
Such distributions have not been measure yet, so that the
prediction will be tested by the further measurement of the Belle
group or even a more accurate measurement at the proposed Super-B
factory.

As aforementioned there are several puzzles for the dipion emission of $\Upsilon(4S)$.
The anomalous ratio of ${\mathcal{R}}=\frac{\Gamma(\Upsilon(4S)\to \Upsilon(2S)\pi^+\pi^-)}{\Upsilon(4S)\to \Gamma(\Upsilon(1S)\pi^+\pi^-)}\approx1.16$;
and the lineshapes $d\Gamma(\Upsilon(4S)\to \Upsilon(2S,1S)\pi^+\pi^-)/dm_{\pi^+\pi^-}$ cannot be explained by
the QCD multipole expansion theory, even though the theory works well for obtaining dipion emission rates of lower $\Upsilon$ states. Noticing the mass
of $\Upsilon(4S)$ is just above the production threshold of $B\bar B$ channel, it is natural to conjecture that on-shell $B\bar B$ is an intermediate state and the sequential
process  $\Upsilon(4S)\to B\bar B\to \Upsilon(nS)+S\to \Upsilon(nS)+\pi^+\pi^-$ may play a crucial role. Namely this sequential process may have a sizable
amplitude and interfere with the direct dipion emission from $\Upsilon(4S)$. The sum of the the two contributions (the two amplitude may have a relative phase,
which is fixed by fitting data in this work) indeed results in a satisfactory explanation to the data. From Fig. (2), one can note that
our model satisfactorily reproduce the lineshapes given by experimental measurements. Moreover, in Fig. (3), our prediction on the angular distribution
$d\Gamma/d\cos\theta$ will be testified by the further measurements of the Belle-II group and/or maybe, the proposed Super-B factory.

This result confirms our conjecture that the long-distance effect, which can be treated as a final state interaction indeed plays a crucial role. Basically
the direct dipion emission occurs via the OZI suppressed process, whereas, the open-bottom process $\Upsilon(4S)\to B\bar B$ does not suffer from
the suppression. Thus even though the final state interaction is realized via hadronic loops, its amplitude may be comparable with the OZI suppressed direct dipion
emission, so that the interference is important. For lower $\Upsilon(nS)$ $(n\leq 3)$, $B\bar B$ cannot be on-shell, the absorptive part of
the hadronic loop does not exist and the contribution may be negligible.

As a conclusion, the puzzles of the dipion emission of $\Upsilon(4S)$ can be satisfactorily removed by taking into account the final state interactions.

\vfill

\section*{Acknowledgement} We would like to thank Feng-Kun Guo for useful discussion. This
project is supported by the National Natural Science Foundation of
China under Grants Nos. 11175073, 11005129, 11035006, 11047606, 11075079,
the Ministry of Education of China (FANEDD under Grant No. 200924,
DPFIHE under Grant No. 20090211120029, NCET, the Fundamental
Research Funds for the Central Universities), and the West
Doctoral Project of Chinese Academy of Sciences.

\end{document}